%% file: template.tex
\documentclass[cameraready]{Interspeech}
\usepackage{tabularx}  % 핵심: 자동 줄바꿈 표 기능
\usepackage{booktabs}  % 가로선 예쁘게 그리기 (toprule, midrule 등)
\usepackage{multirow}  % 행 합치기 기능
\usepackage{subcaption}

\title{A Hierarchical Feature Engineering Framework for Automated Classification of Phonotraumatic and Non-Phonotraumatic Vocal Hyperfunction}

\author[affiliation={1,2,3}, orcid=0000-0003-0111-300X]{June-Woo}{Kim}
\author[affiliation={4}, orcid=0009-0004-9533-4595]{Kangwook}{Jang}
\author[affiliation={4}]{Minu}{Kim}
\author[affiliation={3,5}, orcid=0000-0003-2389-7183]{Hyunju}{Lee$^\dagger$}
\address{
    $^1$ Department of Electronic Engineering, Wonkwang University, Republic of Korea \\
    $^2$ AI Convergence Research Institute, Wonkwang University, Republic of Korea \\
    $^3$ GIST InnoCORE AI-Nano Convergence Institute for Early Detection of Neurodegenerative Diseases, Gwangju Institute of Science and Technology, Republic of Korea \\
    $^4$ School of Electrical Engineering, KAIST, Republic of Korea \\
    $^5$ Department of AI Convergence, Gwangju Institute of Science and Technology, Republic of Korea
}

\email{kaen2891@wku.ac.kr, hyunjulee@gist.ac.kr}


\usepackage{comment}

\begin{document}

\maketitle
\renewcommand{\thefootnote}{$\dagger$}
\footnotetext{Corresponding author.}

% the abstract here must exactly match the abstract entered into the paper submission system
\begin{abstract}
Ambulatory neck-surface acceleration enables non-invasive monitoring of vocal hyperfunction, yet robust biomarkers for its subtypes remain limited. This study investigates the NeckVibe Challenge dataset to distinguish phonotraumatic (PVH) and non-phonotraumatic (NPVH) from healthy controls. We propose a hierarchical feature engineering framework comprising: $(i)$ static, $(ii)$ dynamic, $(iii)$ ratio-based, $(iv)$ coupling features capturing source filter interactions. While univariate statistical analysis shows strong separability for PVH but limited significance for NPVH, our machine learning pipeline, tailored for high-dimensional feature integration, identifies that coupling features are crucial for both tasks. We achieve an AUC of 0.891 for PVH and 0.728 for NPVH, suggesting that while PVH is near-linearly separable, NPVH discrimination benefits from modeling non-linear feature interactions.
%evaluated through 10-fold stratified group cross-validation, 
%We analyze the NeckVibe Challenge to distinguish  (NPVH) from controls using subject-level features. To this end, acoustic and IBIF-derived aerodynamic measures are summarized as (i) static distribution statistics, (ii) dynamic delta and trend descriptors, (iii) normalized ratios, and (iv) physiologically motivated coupling ratios capturing source filter and stability effort relations. Welch tests with FDR show strong separability for PVH but none for NPVH. With 10-fold subject-level cross-validation and RFECV, coupling features yield the best performance, reaching AUC 0.891 for PVH and improving NPVH discrimination to AUC 0.728 (F1 0.747). Results suggest PVH is near-linearly separable, whereas NPVH benefits from nonlinear interaction modeling.
\end{abstract}

\input{1-Introduction}
\input{2-Preliminaries}
\input{3-Method}
\input{4-Experiments}
\input{5-Discussion}
%\input{6-Conclusion}

%\pagebreak
\bibliographystyle{IEEEtran}
\bibliography{mybib}

\end{document}

%% file: 1-Introduction.tex
\section{Introduction}
%Vocal hyperfunction is a broad clinical construct describing excessive or imbalanced laryngeal and supralaryngeal muscle activity during phonation, often associated with voice fatigue, reduced vocal efficiency, and, in some cases, tissue trauma. In clinical practice, hyperfunctional patterns span both phonotraumatic vocal hyperfunction (PVH), which is more directly linked to vocal fold lesions, and non-phonotraumatic vocal hyperfunction (NPVH), which can present with subtler perceptual and physiological manifestations. Because hyperfunctional behaviors frequently vary across tasks, environments, and time, assessments confined to clinic visits can miss important fluctuations that occur in daily life. This motivates ambulatory monitoring approaches that can capture ecologically valid voice use patterns over extended periods.
Vocal hyperfunction (VH) is a prevalent condition characterized by chronic voice misuse, leading to disorders categorized as phonotraumatic VH (PVH, e.g., nodules) and non-phonotraumatic VH (NPVH, e.g., muscle tension dysphonia)~\cite{hillman2020updated, kridgen2021patient}. While clinical diagnosis is standard, ambulatory monitoring using neck-surface acceleration (ACC) has emerged as a powerful tool for capturing daily vocal behavior in naturalistic settings~\cite{mehta2012mobile, mehta2015using, van2020differences, cortes2022ambulatory}. However, distinguishing PVH subtypes from healthy controls remains challenging due to the high variability of acoustic and aerodynamic measures in daily life~\cite{van2021differences}.

%Neck-surface acceleration sensing offers a practical route to ambulatory monitoring by measuring vibration at the neck surface with minimal intrusion. Compared to microphone recordings, this modality is less sensitive to environmental acoustics and can be deployed continuously in real-world settings. However, despite the promise of long-term monitoring, reliable biomarkers for vocal hyperfunction remain uncertain. In particular, it is not always clear which features generalize across subjects and recording days, and which representations best capture clinically relevant differences between PVH, NPVH, and controls. To this end, we use the NeckVibe Challenge dataset, which provides a standardized testbed for this problem by defining two classification tasks, PVH versus control and NPVH versus control, using ambulatory neck-surface vibration recordings.
Recent studies have utilized various features derived from ACC signals, such as aerodynamic measures estimated via Impedance-based Inverse Filtering (IBIF)~\cite{zanartu2013subglottal} to quantify glottal airflow~\cite{van2020differences, cortes2022ambulatory, van2021differences, mehta2019difference, morales2023glottal}. While these measures provide insights into vocal fold physiology, most approaches rely on simple time-averages of individual features, potentially overlooking the dynamic and interdependent nature of speech production~\cite{ghassemi2014learning}. In particular, the \emph{physiological coupling} between different vocal parameters, such as the relationship between aerodynamic effort and acoustic output is often neglected in automated classification pipelines~\cite{morales2023glottal, fryd2016estimating}.

Here, we investigate the NeckVibe Challenge dataset to improve PVH and NPVH classification. Our contributions are: %three-fold:
\begin{itemize}
    \item We propose a hierarchical feature engineering framework categorized into static distributions, dynamic trend descriptors, ratio-based proportions, and physiologically motivated coupling features to capture the complex mechanism of vocal hyperfunction.
    \item We perform rigorous univariate statistical analysis using Welch's t-test with Benjamini-Hochberg false discovery rate (FDR)~\cite{benjamini1995controlling} correction to identify features that distinguish PVH and NPVH, providing clinical interpretability.
    \item We employ a recursive feature elimination cross-validation (RFECV) pipeline to demonstrate that while PVH is near-linearly separable, NPVH classification benefits significantly from capturing non-linear feature interactions through our proposed feature set.
    \item We validate our framework using a stratified 10-fold group cross-validation on the given NeckVibe Challenge dataset. The model achieves an AUC of 0.891 for Task 1 (PVH vs. Control) and 0.728 for Task 2 (NPVH vs. Control), providing a competitive baseline and demonstrating the practical utility of our hierarchical features for biomarker identification in vocal hyperfunction.
\end{itemize}
%%

%We validate this hypothesis using subject-level 10-fold cross-validation and RFECV feature selection to assess which feature groups generalize best. In addition, we perform univariate analysis using Welch tests with FDR control to contrast separability patterns across tasks. Our results show a clear contrast between PVH and NPVH: PVH exhibits strong separability and achieves high AUC, while NPVH shows limited univariate separation but benefits more from interaction-oriented representations. Overall, coupling features yield the best performance, reaching AUC 0.891 for PVH and improving NPVH discrimination to AUC 0.728 with F1 0.747. These findings suggest that PVH is near-linearly separable at the subject level, whereas NPVH may require modeling nonlinear interactions among production-related cues.

%% file: 2-Preliminaries.tex
\input{Tables/table1}

\section{NeckVibe Challenge Dataset}
%The NeckVibe Challenge provides ambulatory neck surface acceleration-derived measurements for subject-level voice disorder detection. The challenge defines two binary classification tasks: $(i)$ PVH versus PVH Control and $(ii)$ non-PVH (NPVH) versus NPVH Control. The released data are frame-level time series features computed at approximately 50 ms resolution, organized by monitoring day (stored as separate MATLAB files) with multiple days available per subject. %Because the evaluation on the test set requires a single prediction per subject, all analyses in this work use subject-level representations obtained by aggregating information across available monitoring days. 

The NeckVibe Challenge provides ambulatory neck surface acceleration-derived measurements for subject-level voice disorder detection. The challenge defines two binary classification tasks: $(i)$ PVH versus PVH Control and $(ii)$ NPVH versus NPVH Control. The dataset comprises 3,278 samples from 468 subjects. Specifically, PVH includes 1,151 samples from 171 subjects, and PVH Control includes 992 samples from 136 subjects. NPVH includes 637 samples from 93 subjects, and NPVH Control includes 498 samples from 68 subjects. IBIF-based aerodynamic variables are partially missing for a subset of subjects and samples: 9 subjects have missing IBIF measurements in at least one recording (PVH: 6, NPVH: 2, NPVH Control: 1), corresponding to 55 samples with missing IBIF (PVH: 37, NPVH: 11, NPVH Control: 7), while PVH Control has no missing IBIF. %To meet the challenge requirement of subject-level prediction, we aggregated all available monitoring days for each subject.

As shown in Table~\ref{tab1:core_features}, the dataset comprised of 14 core physical quantities, consisting of six acoustic measures %(H1H2all, LHratioall, cppall, dBcms2, level, spectralTiltall) 
and eight IBIF-based aerodynamic measures. %(IBIF\_h1h2, IBIF\_hrf, IBIF\_ mfdr, IBIF\_acflow, IBIF\_sq, IBIF\_oq, IBIF\_naq, IBIF\_cq). 
Each frame includes a voiced indicator, and we restrict feature computation to voiced frames to focus on phonatory segments. 
For statistical analyses, observations containing imputed or missing values were excluded to avoid bias in group-level differences. In contrast, for machine learning (ML) experiments, missing values were processed using median imputation within each training fold. 
%No external data was used.

%% file: Tables/table1.tex
\begin{table*}[t]
\centering
\caption{Core acoustic and aerodynamic (IBIF-derived) features extracted from neck-surface acceleration signals.}
\label{tab1:core_features}
\footnotesize
%\resizebox{\linewidth}{!}{
\begin{tabular}{lll}
\hline
\textbf{Category} & \textbf{Variable} & \textbf{Description (Unit)} \\
\hline
\multirow{6}{*}{Acoustic} 
& H1H2all & Harmonic magnitude difference between H1 and H2; reflects glottal configuration (dB) \\
& LHratioall & Low/high frequency power ratio (70--2000 Hz vs. 2000--3730 Hz); spectral balance (dB) \\
& cppall & Cepstral peak prominence; periodicity and dysphonia measure (dB) \\
& dBcms2 & RMS neck-surface acceleration amplitude (dB re cm/s$^2$) \\
& level & Log-scaled RMS neck-surface acceleration amplitude (dB re cm/s$^2$) \\
& spectralTiltall & Spectral tilt; regression slope across first eight harmonics (dB/oct) \\
\hline
\multirow{8}{*}{Aerodynamic (IBIF)} 
& IBIF\_h1h2 & Harmonic magnitude difference H1--H2 from estimated airflow (dB) \\
& IBIF\_hrf & Harmonic richness factor; summed harmonics relative to first harmonic (dB) \\
& IBIF\_mfdr & Maximum flow declination rate; peak negative airflow derivative (L/s$^2$) \\
& IBIF\_acflow & Peak-to-peak glottal airflow amplitude (mL/s) \\
& IBIF\_sq & Speed quotient (ratio of opening to closing phase duration) (unitless) \\
& IBIF\_oq & Open quotient; proportion of glottal cycle open (unitless) \\
& IBIF\_naq & Normalized amplitude quotient; ACFL/MFDR normalized by period (unitless) \\
& IBIF\_cq & Closing quotient; proportion of glottal cycle in closing phase (unitless) \\
\hline
\vspace{-5mm}
\end{tabular}
%}
\end{table*}

%% file: 3-Method.tex
\section{Methods}
\subsection{Feature Construction}
To capture the complex mechanisms of VH, we constructed a hierarchical feature set focusing on static, dynamic, ratio-based, and coupling properties. We include the \emph{vocal dose percentage} in all configurations, defined as the proportion of voiced frames relative to total recording time. All frame-level measures were aggregated to subject-level representations.

% \subsubsection{Static Features} 
\vspace{3pt} \noindent \textbf{Static Features.} For each core acoustic and IBIF variable, seven statistical descriptors were computed to characterize central tendency, variability, and distributional shape:\,mean, standard deviation (SD), 5th percentile ($P_5$), 95th percentile ($P_{95}$), skewness, kurtosis, interquartile range. The total number of static features is $14 \times 7 + 1 = 99$, where 14 denotes the number of base features with the vocal dose percentage.

% \subsubsection{Dynamic Features} 
\vspace{3pt} \noindent \textbf{Dynamic Features.} Temporal dynamics were modeled using first-order ($\Delta$) and second-order ($\Delta\Delta$) differences statistics, including delta mean, delta SD, upper-tail delta magnitude ($P_{95}$), mean absolute delta, delta–delta SD, and linear trend (slope). 
The dynamic feature configuration consists of $14 \times 7$ additional descriptors combined with the static features, resulting in a total of $197$ features.

% \subsubsection{Ratio-based Features}
\vspace{3pt} \noindent \textbf{Ratio-based Features.} To obtain normalized representations, four additional relative variability measures were computed for each base feature: $\Delta$SD/mean, $\Delta$SD/SD, $|$slope$|$/mean, and $|\Delta|$/mean.
These descriptors quantify temporal variability relative to absolute level or dispersion. 
The ratio-based configuration therefore adds $14 \times 4$ features to the dynamic feature set, resulting in a total of $14 \times 4 + 197 = 253$ features.

% \subsubsection{Coupling Features} 
\vspace{3pt} \noindent \textbf{Coupling Features.} Six physiologically motivated~\cite{morales2023glottal, fryd2016estimating} interaction terms were introduced to capture relationships between acoustic and aerodynamic properties. 
These include source-filter coupling (cppall/spectralTiltall, cppall/H1H2all), stability-effort coupling (cppall/$\Delta$SD, cppall/abs$\Delta$), and aerodynamic intra-feature coupling (IBIF\_naq/$\Delta$SD, IBIF\_oq/abs$\Delta$). 
The coupling configuration extends the dynamic feature set by six interaction terms, resulting in a total of $197 + 6 = 203$ features.

\subsection{Univariate Statistical Analysis}
Statistical comparisons were performed for each task (Task\,1 vs. Task\,2). Group differences were assessed independently using \textbf{\emph{Welch's t-tests}}, which do not assume equal variances or sample sizes. The magnitude of group differences was quantified using \textbf{\emph{Cohen's d}} effect size. To control for Type I errors in high-dimensional testing, $p$-values were adjusted using the \textbf{\emph{Benjamini--Hochberg FDR}} procedure within each feature set.

\begin{figure*}[t]
\centering

\begin{subfigure}[t]{0.48\linewidth}
    \centering
    \includegraphics[width=\linewidth]{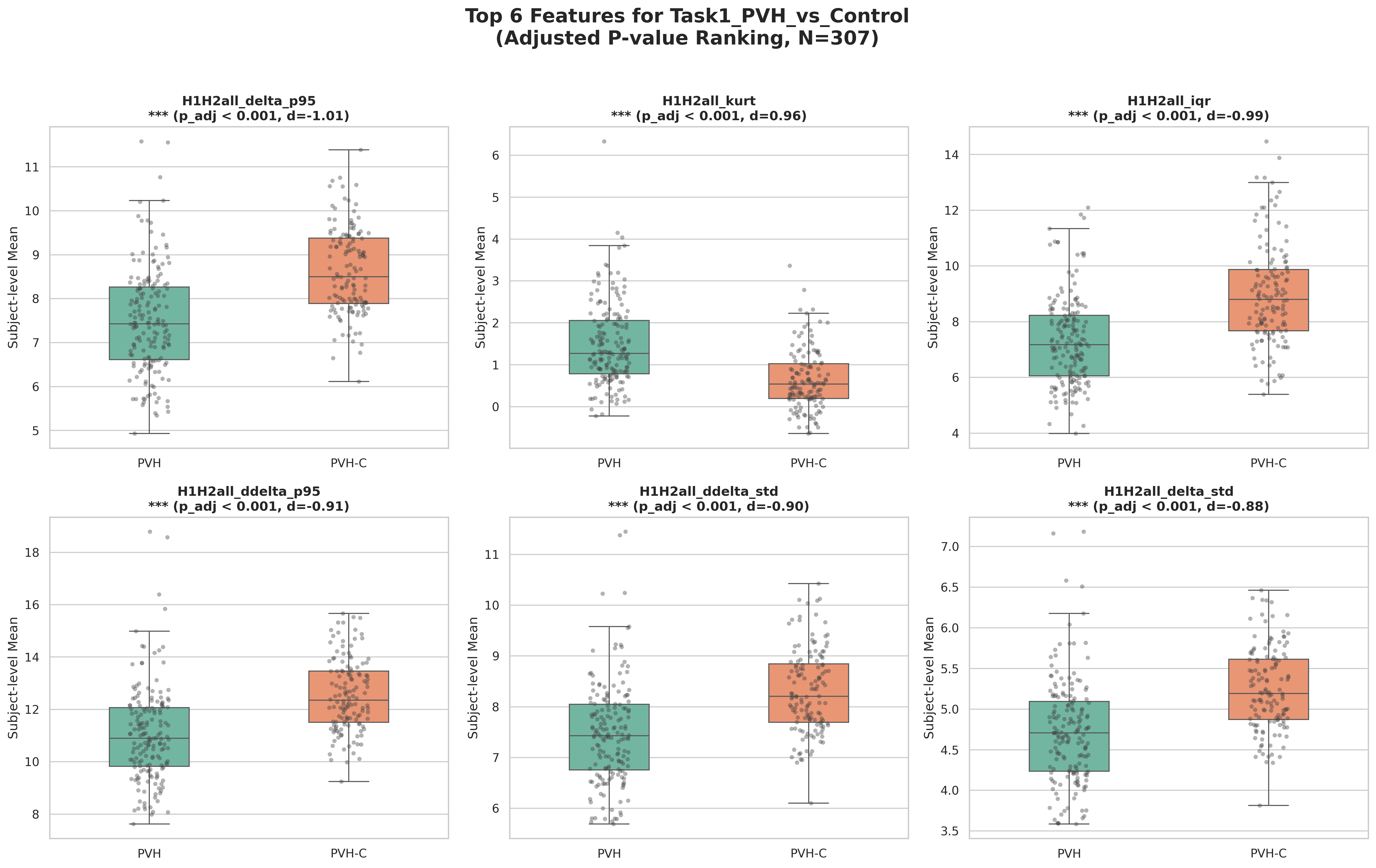}
    \caption{Task~1 (PVH vs. Control). Top six subject-level features ranked by FDR-adjusted $p$-value. All displayed features remain statistically significant after multiple comparison correction ($p_{adj}<0.001$), with large effect sizes ($|d| \approx 0.9$--1.0).}
    \label{fig:task1}
\end{subfigure}
\hfill
\begin{subfigure}[t]{0.48\linewidth}
    \centering
    \includegraphics[width=\linewidth]{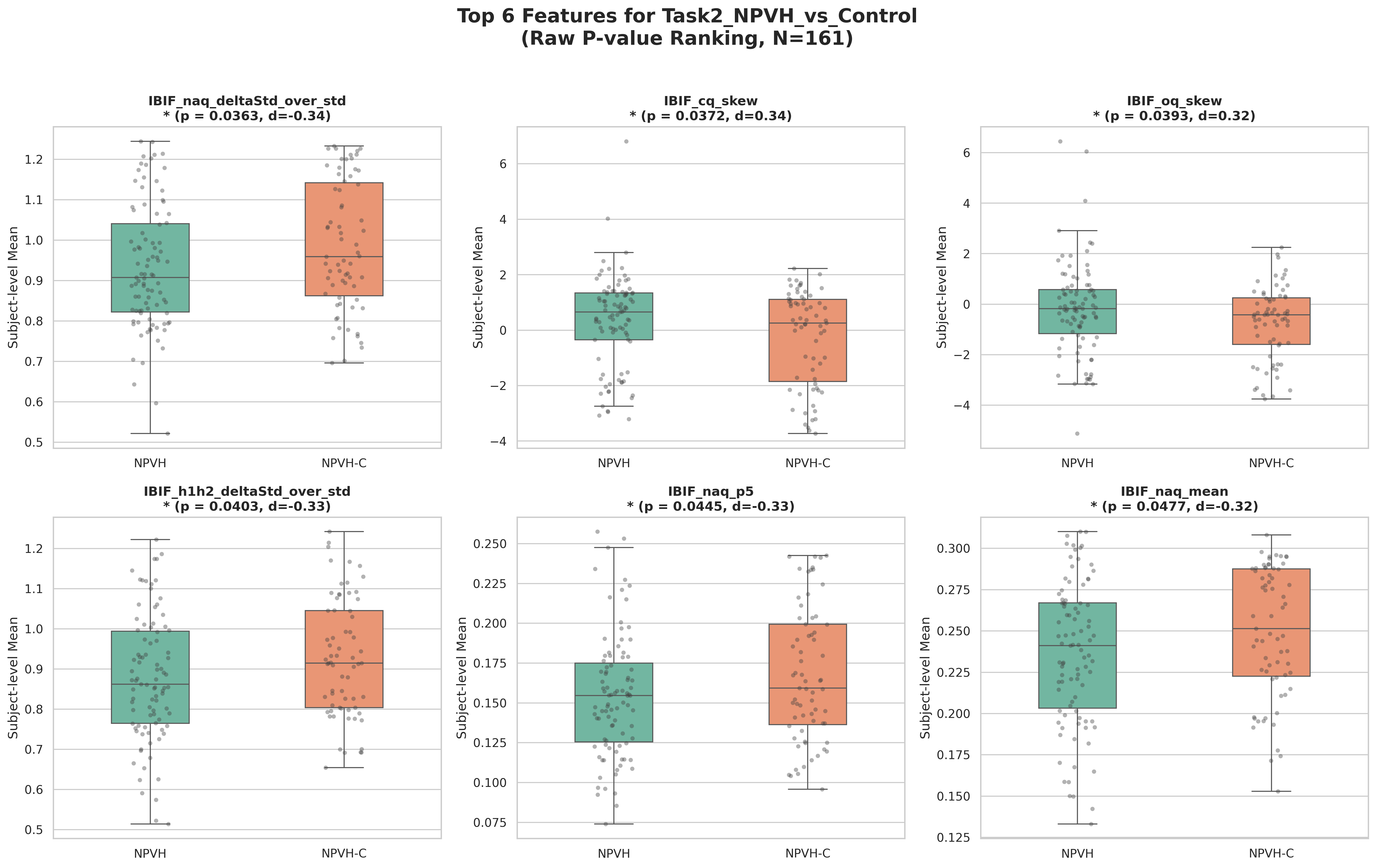}
    \caption{Task~2 (NPVH vs. Control). Top six subject-level features ranked by raw $p$-value. No features survived FDR correction; features with nominal significance ($p<0.05$) are shown to illustrate exploratory trends. Effect sizes are smaller than Task 1 ($|d| \approx 0.3$).}
    \label{fig:task2}
    \vspace{-5mm}
\end{subfigure}

\caption{Comparison of statistical features for PVH (Task 1) and NPVH (Task 2). While PVH exhibits robust group separation characterized by large effect sizes across multiple dynamic and higher-order statistics, NPVH shows only weak and inconsistent differences at the daily summary level, even among the most discriminative features.}
\vspace{-5mm}
\label{fig:stat_comparison}
\end{figure*}

\subsection{Machine Learning Analysis}
%To capture multivariate interactions, we evaluated five classifiers: Logistic Regression, SVM (RBF), Random Forest, XGBoost, and LightGBM.
\noindent \textbf{Cross-Validation Strategy by Subject-Level.} We conducted supervised classification experiments at the subject level to determine whether multivariate modeling could capture interaction patterns beyond univariate statistical differences.
%To examine whether multivariate modeling can capture interaction patterns beyond univariate statistical differences, we conducted supervised classification experiments at the subject level. 
All experiments were performed separately for PVH vs. PVH Control (Task 1) and NPVH vs. NPVH Control (Task 2), using the same subject-level feature representations described in the previous section.
We implemented a stratified 10-fold CV scheme using subject IDs as grouping variables which guarantees that all acoustic recordings from a specific individual are strictly confined to either the training or validation set within each fold.

%\subsubsection{Feature Configurations}
%Four feature configurations were evaluated: $(i)$ static statistics (99 features), $(ii)$ dynamic features (197 features), $(iii)$ ratio-based features (253 features), and $(iv)$ inclusion of coupling features (203 features). Each configuration was assessed independently to analyze the incremental contribution of dynamic and relational representations.

% \subsubsection{Cross-Validation Strategy by Subject-Level}
% To avoid subject-level data leakage, we employed a 10-fold StratifiedGroupKFold cross-validation scheme, where subject IDs were treated as grouping variables. This can verify that all recordings from a given subject were confined to either the training or validation set within each fold.

% \subsubsection{Automatic Feature Selection by ML Models}
% To reduce redundancy and improve model stability in the high-dimensional feature space, RFECV was applied using a tree-based estimator (XGBoost) within the training folds. The selected feature subset was then used consistently across all evaluated classifiers. We also tried the greedy search algorithm, however, the performance is not promising.

\vspace{3pt} \noindent \textbf{Feature Selection by ML Models.} To further address high-dimensional feature space redundancy and enhance model stability, we performed automatic feature selection exclusively within these designated training folds motivated by work in schizophrenia screening~\cite{jang2025acoustic}. Specifically, RFECV utilizing an XGBoost tree-based estimator was applied. While we also evaluated a preliminary greedy search algorithm, it yielded suboptimal classification performance. Consequently, the optimal feature subset identified via RFECV was utilized consistently across all subsequent classifier evaluations.

% \subsubsection{Classification Models}
\vspace{3pt} \noindent \textbf{Classification Models.} Five ML classifiers were evaluated: logistic regression~\cite{cox1958regression}, support vector machine (SVM) with Gaussian  kernel~\cite{cortes1995support}, random forest~\cite{breiman2001random}, XGBoost~\cite{chen2016xgboost}, lightGBM~\cite{ke2017lightgbm}. 
The experimental design was intended to compare statistical feature representations under controlled conditions. 
The primary objective was to assess how different feature structures influence classification performance, rather than to optimize predictive accuracy through extensive hyperparameter search. Accordingly, fixed hyperparameter configurations were used and applied across folds and feature sets.

% \subsubsection{Model Interpretation}
\vspace{3pt} \noindent \textbf{Model Interpretations.} To analyze feature contribution patterns and model behavior, SHapley Additive exPlanations (SHAP)~\cite{lundberg2017unified} were computed for the best-performing model in each task using a model-agnostic explainer. SHAP values were aggregated across validation folds to examine whether predictions were driven by a small set of dominant features or by distributed multivariate interactions.

\input{Tables/table2}

%% file: Tables/table2.tex
\begin{table}[t]
\centering
\caption{Summary of statistically significant features across different feature configurations for each task. \textbf{Task~1 Sig.} denotes the number of features with FDR-adjusted $p$-values that are significant ($p_{adj}<0.05$), whereas \textbf{Task~2 Sig.} indicates the number of features with nominal significance ($p<0.05$), reported due to the absence of FDR-significant features for Task~2.}
\label{tab2:stat_summary}

\resizebox{\linewidth}{!}{%eee
\begin{tabular}{lccc}
\hline
\textbf{Feature Configuration} & \textbf{\# Features} & \textbf{Task~1 Sig.} & \textbf{Task~2 Sig.} \\
\hline
Static features & 99  & 65  & 4  \\
Dynamic features & 197 & 123 & 4 \\
Ratio-based features & 253 & 130 & 6 \\
Coupling features & 203 & 119 & 4 \\
\hline
\end{tabular}%
}
\vspace{-5mm}
\end{table}

%% file: 4-Experiments.tex
\section{Results}
\subsection{Statistical Results}
A summary of statistically significant features across different feature configurations is provided in Table~\ref{tab2:stat_summary}. 
For Task~1, a substantial number of features remained statistically significant after FDR correction. 
The number of significant features increased as dynamic and normalized descriptors were incorporated. As illustrated in Fig.~\ref{fig:task1}, the most discriminative features for PVH primarily involved dynamic and higher-order distributional descriptors, exhibiting large effect sizes and consistent group separation.
In contrast, for Task~2, no features survived FDR correction under any configuration. 
Therefore, features with nominal significance ($p<0.05$) were reported to indicate potential exploratory trends. 
These findings reveal a notable asymmetry between PVH and NPVH. While PVH can be robustly characterized using univariate statistics, NPVH exhibits minimal separability under the same statistical framework.
%A summary of statistically significant features across different feature configurations is provided in Table~\ref{tab2:stat_summary}. For Task~1 (PVH vs. Control), a substantial proportion of features remained statistically significant after FDR correction, with the number of significant features increasing as additional dynamic and normalized statistics were introduced. As illustrated in Fig.~\ref{fig:task1}, the most discriminative features for PVH primarily involved dynamic and higher-order distributional descriptors, exhibiting large effect sizes and consistent group separation.

%In contrast, for Task~2 (NPVH vs. Control), no features reached the FDR-adjusted significance threshold across any feature configuration. Consequently, features with nominal significance ($p<0.05$) were reported to highlight potential exploratory trends. As shown in Fig.~\ref{fig:task2}, the top-ranked features for Task~2 were predominantly derived from aerodynamic IBIF measures and exhibited relatively small effect sizes. Notably, the number of nominally significant features remained limited even after incorporating dynamic, ratio-based, and coupling-based features.

\input{Tables/table3}

%Overall, these findings reveal a notable asymmetry between PVH and NPVH. While PVH can be robustly characterized using univariate subject-level statistics, NPVH exhibits minimal separability under the same statistical framework. %This systematic limitation motivates the subsequent use of multivariate and machine learning-based models to capture subtle interaction patterns that are not detectable through univariate statistical analysis alone.

\subsection{Machine Learning Results}

Table~\ref{tab3:ml_main} shows the classification performance across feature configurations, reported as mean $\pm$ SD over stratified 10-fold CV.

%\subsubsection{Classification Performance}

\noindent \textbf{Task 1 (PVH vs. PVH-Control).}  
Task 1 shows consistently high performance across feature configurations. Static features already yield strong discrimination (AUC = 0.851), and adding dynamic (0.869) and ratio features (0.885) provides incremental gains. The best result is achieved with coupling features using logistic regression (AUC = 0.891 $\pm$ 0.04), while the highest F1 is obtained with ratio features (F1 = 0.838).
%All feature configurations in Table~\ref{tab2:stat_summary} yielded strong discrimination performance. Static statistics alone achieved an AUC of 0.851, indicating that summary descriptors already contain substantial discriminative information. Incorporating dynamic features improved the AUC to 0.869, while ratio-based representations further increased it to 0.885. The highest AUC was obtained using coupling features (0.891 $\pm$ 0.04), suggesting that relational modeling between excitation, spectral, and stability measures provides additional discriminative structure. Improvements were also reflected in the F1-score, with the best F1 achieved using ratio-based features (0.838).

\noindent \textbf{Task 2 (NPVH vs. Control).} 
Task 2 is substantially more challenging. Static features perform near chance (AUC = 0.556), whereas dynamic features improve AUC to 0.682; ratio features do not consistently help (AUC = 0.608). The strongest performance is achieved with LightGBM-based coupling features (AUC = 0.728 $\pm$ 0.10, F1 = 0.747), although overall separability remains lower than in Task 1.
%In contrast, classification performance for Task 2 was substantially lower across all configurations. Static features resulted in near-chance discrimination (AUC = 0.556). The inclusion of dynamic features increased AUC to 0.682, while ratio-based features did not provide consistent gains (AUC = 0.608). The most notable improvement was observed with coupling features, which achieved an AUC of 0.728 $\pm$ 0.10 and the highest F1-score (0.747). Nevertheless, overall separability remained lower bound compared to Task 1.

\subsection{Effect of Feature Representations}
The results demonstrate a clear asymmetry between the two tasks. For Task 1, performance increases progressively as feature representations become more expressive, indicating that PVH-related differences manifest across static, dynamic, and relational dimensions. In contrast, Task 2 exhibits limited separability under static and ratio-based representations, and only modest improvement when coupling features are introduced. 

These findings are consistent with the univariate statistical analysis, where PVH showed robust statistical differences across multiple descriptors, whereas NPVH exhibited minimal separability at the daily summary level. Multivariate modeling partially mitigates this limitation for NPVH but does not close the performance gap observed in Task 1.

\subsection{Feature Contribution Analysis}
\begin{figure}[t]
\centering
\includegraphics[width=\columnwidth]{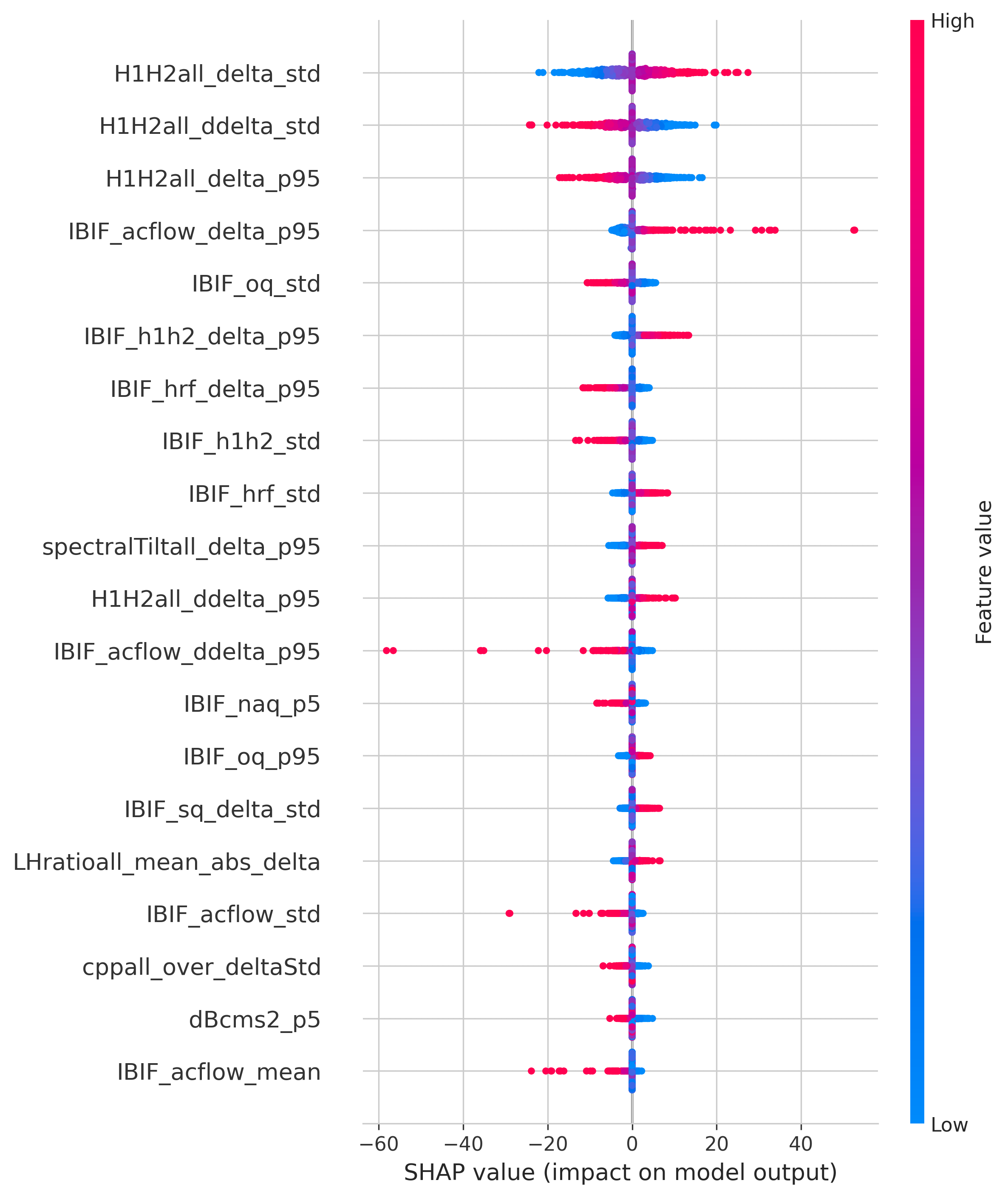}
\caption{Task 1 SHAP summary (Logistic Regression).}
\vspace{-5mm}
\label{fig3:shap_task1}
\end{figure}

\begin{figure}[t]
\centering
\includegraphics[width=\columnwidth]{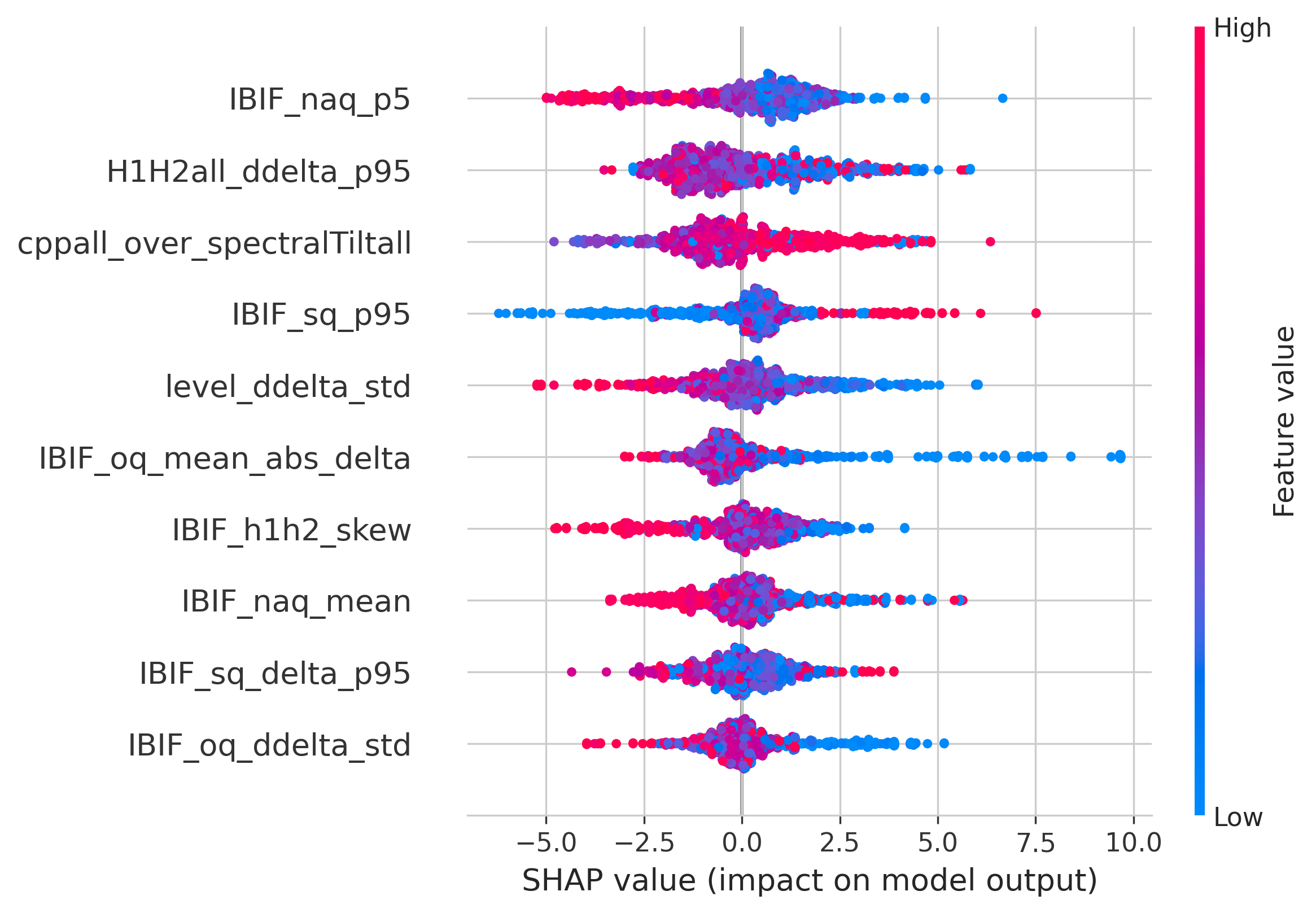}
\caption{Task 2 SHAP summary (LightGBM).}
\vspace{-3mm}
\label{fig4:shap_task2}
\end{figure}

SHAP summary plots for the best-performing model in each task are shown in Fig.~\ref{fig3:shap_task1} and Fig.~\ref{fig4:shap_task2}. For Task 1, feature contributions are distributed across both static and dynamic descriptors. Several delta-based statistics (e.g., harmonic variability and upper-tail delta measures) exhibit consistent directional influence on model output. The SHAP value distribution is relatively symmetric and stable, indicating that classification is supported by multiple moderately contributing features rather than a single dominant predictor.

In contrast, Task 2 shows a more concentrated contribution pattern. Dynamic variability measures, particularly higher-order delta statistics of aerodynamic features (e.g., naq), exhibit stronger localized SHAP magnitudes. The broader spread and higher variance of SHAP values across folds suggest a less stable discriminative structure compared to Task 1.

\subsection{Comparison with Univariate Analysis}
The multivariate modeling results are consistent with the statistical findings reported earlier. Task 1 demonstrated widespread univariate separability across static, dynamic, and relational descriptors, which translated into robust classification performance across models. Task 2, however, exhibited minimal univariate significance and correspondingly limited classification accuracy. Although nonlinear models partially improved discrimination in Task 2, the overall performance gap between the two tasks remained substantial. These findings indicate that PVH-related ambulatory voice patterns exhibit strong and distributed multivariate structure, whereas NPVH-related patterns show weaker and more interaction-dependent characteristics.

\subsection{Performance on the Challenge Held-Out Test Set}
Finally, we evaluated our best model on the unseen NeckVibe Challenge test set. %Performance metrics are summarized in Table~\ref{tab4:results}. 
For Task 1, we achieved a high discriminative performance with an \textbf{AUC of 0.917}. This high performance demonstrates the efficacy of the proposed hierarchical feature framework in capturing the complex physiological patterns of phonotraumatic vocal hyperfunction. %This result confirms that phonotraumatic lesions leave distinct, stable signatures in ambulatory neck-acceleration signals, which our hierarchical features—especially the static and dynamic descriptors—successfully captured.
In contrast, we obtained an AUC of 0.579 in Task 2. This suggests that the pathological contrast in NPVH is insufficiently distinct for multivariate modeling. %Given that no features survived FDR correction in our univariate analysis, the minimal separation between groups likely prevented the classifiers from identifying generalizable patterns, leading to the observed decline in performance.

%% file: Tables/table3.tex
\begin{table*}[t]
\centering
\caption{Performance (mean $\pm$ standard deviation) over 10-fold cross-validation. Best results for each task are highlighted in bold.}
\label{tab3:ml_main}
\small
\resizebox{\linewidth}{!}{%
\begin{tabular}{lccccc|ccccc}
\hline
\multirow{2}{*}{\textbf{Method}} 
& \multicolumn{5}{c|}{\textbf{Task 1 (PVH vs PVH-Control)}} 
& \multicolumn{5}{c}{\textbf{Task 2 (NPVH vs NPVH-Control)}} \\
\cline{2-11}
& AUC & Acc & Prec & Rec & F1 
& AUC & Acc & Prec & Rec & F1 \\
\hline

Baseline & 0.820~\cite{van2020differences, cortes2018ambulatory} & - & - & - & - & \textbf{0.780}~\cite{van2021differences} & - & - & - & - \\

\hline

Static 
& 0.851 $\pm$ 0.05 
& 0.811 $\pm$ 0.04 
& 0.851 $\pm$ 0.08 
& 0.818 $\pm$ 0.07 
& 0.828 $\pm$ 0.03 
& 0.556 $\pm$ 0.21 
& 0.552 $\pm$ 0.16 
& 0.594 $\pm$ 0.14 
& 0.689 $\pm$ 0.21 
& 0.631 $\pm$ 0.16 \\

Dynamic 
& 0.869 $\pm$ 0.04 
& 0.821 $\pm$ 0.05 
& \textbf{0.869 $\pm$ 0.04}
& 0.801 $\pm$ 0.09 
& 0.831 $\pm$ 0.05 
& 0.682 $\pm$ 0.10 
& 0.640 $\pm$ 0.05 
& 0.660 $\pm$ 0.05 
& 0.800 $\pm$ 0.13 
& 0.715 $\pm$ 0.06 \\

Ratio 
& 0.885 $\pm$ 0.06 
& \textbf{0.824 $\pm$ 0.08} 
& 0.857 $\pm$ 0.07 
& \textbf{0.825 $\pm$ 0.09} 
& \textbf{0.838 $\pm$ 0.07} 
& 0.608 $\pm$ 0.19 
& 0.589 $\pm$ 0.17 
& 0.628 $\pm$ 0.16 
& 0.670 $\pm$ 0.24 
& 0.639 $\pm$ 0.18 \\

Coupling 
& \textbf{0.891 $\pm$ 0.04} 
& 0.817 $\pm$ 0.05 
& 0.855 $\pm$ 0.04 
& 0.813 $\pm$ 0.11 
& 0.829 $\pm$ 0.06 
& 0.728 $\pm$ 0.10
& \textbf{0.683 $\pm$ 0.05} 
& \textbf{0.693 $\pm$ 0.04} 
& \textbf{0.820 $\pm$ 0.11} 
& \textbf{0.747 $\pm$ 0.05} \\
\hline
\vspace{-7mm}
\end{tabular}
}
\end{table*}

%% file: 5-Discussion.tex
\section{Discussion and Conclusion}
The primary objective of this study was to evaluate the incremental utility of complex feature configurations in characterizing vocal hyperfunction. Our findings reveal a clear divergence in classification performance between Task 1 and Task 2. For Task 1, leveraging coupling features notably enhanced model performance, suggesting that PVH conditions leave distinct physiological signatures in ambulatory voice data. In contrast, the relatively low performance observed in Task 2 reflects a lack of clear discriminatory rationales in the current feature set. In our univariate analysis, no single feature reached statistical significance after FDR correction for NPVH. We hypothesize that NPVH conditions, often characterized by psychological distress or functional imbalances without structural lesions~\cite{hillman2020updated, hillman1989objective}, lead to substantial distributional overlap with the control group, making it difficult for models to identify a robust classification boundary using aerodynamic measures.% the characteristics of NPVH conditions lead to substantial distributional overlap with the control group, making it difficult for machine learning models to identify a robust classification boundary.

%Although the current dataset provides feature-level representations only, future work may investigate self-supervised speech models~\cite{baevski2020wav2vec, hsu2021hubert, chen2022wavlm} trained on raw waveform data, which may better capture subtle temporal dynamics. Such speech foundation models could offer better representation of NPVH conditions by modeling raw temporal sequences directly.
Although the current NeckVibe Challenge dataset provides feature-level representation only, future work can be considered exploring self-supervised speech models~\cite{baevski2020wav2vec, chen2022wavlm} trained on raw waveform data. Such speech foundation models could offer a more nuanced representation of NPVH by modeling raw temporal sequences directly. This approach is particularly promising for NPVH, as its key biomarkers may reside in subtle, non-stationary prosodic variations or micro-tremors in the voice that are often lost during the process of manual feature extraction and temporal averaging~\cite{helou2013intrinsic, kim2026deep}. Investigating these high-dimensional latent representations could potentially explore the \emph{hidden} signatures of non-phonotraumatic laryngeal tension and provide a more robust objective assessment tool. % for clinical use.

\section{Acknowledgement}
This research was supported by the InnoCORE program of the Ministry of Science and ICT(GIST InnoCORE KH0860), and by the Regional Innovation System \& Education(RISE) program through the Jeonbuk RISE Center, funded by the Ministry of Education(MOE) and the Jeonbuk State, Republic of Korea(2026-RISE-13-WKU).
%Brian Impact Foundation, a non-profit organization dedicated to the advancement of science and technology for all.

\section{Generative AI Use Disclosure}
Generative AI (ChatGPT) was used solely for grammar correction and linguistic polishing of this manuscript. The authors have verified all technical content and maintain full accountability for the work.